\input harvmac
\input graphicx
\input color
%
%
%
\message{S-Tables Macro v1.0, ACS, TAMU (RANHELP@VENUS.TAMU.EDU)}
%
%
\newhelp\stablestylehelp{You must choose a style between 0 and 3.}%
\newhelp\stablelinehelp{You should not use special hrules when stretching
a table.}%
\newhelp\stablesmultiplehelp{You have tried to place an S-Table inside another
S-Table.  I would recommend not going on.}%
%
%
\newdimen\stablesthinline
\stablesthinline=0.4pt
\newdimen\stablesthickline
\stablesthickline=1pt
%
%
\newif\ifstablesborderthin
\stablesborderthinfalse
\newif\ifstablesinternalthin
\stablesinternalthintrue
\newif\ifstablesomit
\newif\ifstablemode
\newif\ifstablesright
\stablesrightfalse
%
%
\newdimen\stablesbaselineskip
\newdimen\stableslineskip
\newdimen\stableslineskiplimit
%
%
\newcount\stablesmode
\newcount\stableslines
\newcount\stablestemp
\stablestemp=3
\newcount\stablescount
\stablescount=0
\newcount\stableslinet
\stableslinet=0
%
%
%
\newcount\stablestyle
\stablestyle=0
%
%
\def\stablesleft{\quad\hfil}%
\def\stablesright{\hfil\quad}%
%
%
%
%
\newcount\stablestrutsize
\newbox\stablestrutbox
\setbox\stablestrutbox=\hbox{\vrule height10pt depth5pt width0pt}
\def\stablestrut{\relax\ifmmode%
                         \copy\stablestrutbox%
                       \else%
                         \unhcopy\stablestrutbox%
                       \fi}%
%
%
\newdimen\stablesborderwidth
\newdimen\stablesinternalwidth
\newdimen\stablesdummy
\newcount\stablesdummyc
\newif\ifstablesin
\stablesinfalse
%
%
\def\begintable{\stablestart%
  \stablemodetrue%
  \stablesadj%
  \halign%
  \stablesdef}%
\def\stablesadj{%
  \ifcase\stablestyle%
    \hbox to \hsize\bgroup\hss\vbox\bgroup%
  \or%
    \hbox to \hsize\bgroup\vbox\bgroup%
  \or%
    \hbox to \hsize\bgroup\hss\vbox\bgroup%
  \or%
    \hbox\bgroup\vbox\bgroup%
  \else%
    \errhelp=\stablestylehelp%
    \errmessage{Invalid style selected, using default}%
    \hbox to \hsize\bgroup\hss\vbox\bgroup%
  \fi}%
\def\stablesend{\egroup%
  \ifcase\stablestyle%
    \hss\egroup%
  \or%
    \hss\egroup%
  \or%
    \egroup%
  \or%
    \egroup%
  \else%
    \hss\egroup%
  \fi}%
\def\stablestart{%
  \ifstablesin%
    \errhelp=\stablesmultiplehelp%
    \errmessage{An S-Table cannot be placed within an S-Table!}%
  \fi
  \global\stablesintrue%
  \global\advance\stablescount by 1%
  \message{<S-Tables Generating Table \number\stablescount}%
  \begingroup%
  \stablestrutsize=\ht\stablestrutbox%
  \advance\stablestrutsize by \dp\stablestrutbox%
  \ifstablesborderthin%
    \stablesborderwidth=\stablesthinline%
  \else%
    \stablesborderwidth=\stablesthickline%
  \fi%
  \ifstablesinternalthin%
    \stablesinternalwidth=\stablesthinline%
  \else%
    \stablesinternalwidth=\stablesthickline%
  \fi%
  \tabskip=0pt%
  \stablesbaselineskip=\baselineskip%
  \stableslineskip=\lineskip%
  \stableslineskiplimit=\lineskiplimit%
  \offinterlineskip%
  \def\borderrule{\vrule width \stablesborderwidth}%
  \def\internalrule{\vrule width \stablesinternalwidth}%
  \def\thinline{\noalign{\hrule height \stablesthinline}}%
  \def\thickline{\noalign{\hrule height \stablesthickline}}%
  \def\trule{\omit\leaders\hrule height \stablesthinline\hfill}%
  \def\ttrule{\omit\leaders\hrule height \stablesthickline\hfill}%
  \def\tttrule##1{\omit\leaders\hrule height ##1\hfill}%
  \def\stablesel{&\omit\global\stablesmode=0%
    \global\advance\stableslines by 1\borderrule\hfil\cr}%
  \def\el{\stablesel&}%
  \def\elt{\stablesel\thinline&}%
  \def\eltt{\stablesel\thickline&}%
  \def\elttt##1{\stablesel\noalign{\hrule height ##1}&}%
  \def\elspec{&\omit\hfil\borderrule\cr\omit\borderrule&%
              \ifstablemode%
              \else%
                \errhelp=\stablelinehelp%
                \errmessage{Special ruling will not display properly}%
              \fi}%
  \def\stmultispan##1{\mscount=##1 \loop\ifnum\mscount>3 \stspan\repeat}%
  \def\stspan{\span\omit \advance\mscount by -1}%
  \def\multicolumn##1{\omit\multiply\stablestemp by ##1%
     \stmultispan{\stablestemp}%
     \advance\stablesmode by ##1%
     \advance\stablesmode by -1%
     \stablestemp=3}%
  \def\multirow##1{\stablesdummyc=##1\parindent=0pt\setbox0\hbox\bgroup%
    \aftergroup\emultirow\let\temp=}
  \def\emultirow{\setbox1\vbox to\stablesdummyc\stablestrutsize%
    {\hsize\wd0\vfil\box0\vfil}%
    \ht1=\ht\stablestrutbox%
    \dp1=\dp\stablestrutbox%
    \box1}%
  \def\stpar##1{\vtop\bgroup\hsize ##1%
     \baselineskip=\stablesbaselineskip%
     \lineskip=\stableslineskip%
     \lineskiplimit=\stableslineskiplimit\bgroup\aftergroup\estpar\let\temp=}%
  \def\estpar{\vskip 6pt\egroup}%
  \def\stparrow##1##2{\stablesdummy=##2%
     \setbox0=\vtop to ##1\stablestrutsize\bgroup%
     \hsize\stablesdummy%
     \baselineskip=\stablesbaselineskip%
     \lineskip=\stableslineskip%
     \lineskiplimit=\stableslineskiplimit%
     \bgroup\vfil\aftergroup\estparrow%
     \let\temp=}%
  \def\estparrow{\vfil\egroup%
     \ht0=\ht\stablestrutbox%
     \dp0=\dp\stablestrutbox%
     \wd0=\stablesdummy%
     \box0}%
  \def\vert{\global\advance\stablesmode by 1&&&}%
  \def\novert{\global\advance\stablesmode by 1&\omit\vrule width 0pt%
         \hfil&&}%
  \def\vt{\global\advance\stablesmode by 1&\omit\vrule width \stablesthinline%
          \hfil&&}%
  \def\vtt{\global\advance\stablesmode by 1&\omit\vrule width \stablesthickline%
          \hfil&&}%
  \def\vttt##1{\global\advance\stablesmode by 1&\omit\vrule width ##1%
          \hfil&&}%
  \def\vtr{\global\advance\stablesmode by 1&\omit\hfil\vrule width%
           \stablesthinline&&}%
  \def\vttr{\global\advance\stablesmode by 1&\omit\hfil\vrule width%
            \stablesthickline&&}%
  \def\vtttr##1{\global\advance\stablesmode by 1&\omit\hfil\vrule width ##1&&}%
  \stableslines=0%
  \stablesomitfalse}
\def\stablesdef{\bgroup\stablestrut\borderrule##\tabskip=0pt plus 1fil%
  &\stablesleft##\stablesright%
  &##\ifstablesright\hfill\fi\internalrule\ifstablesright\else\hfill\fi%
  \tabskip 0pt&&##\hfil\tabskip=0pt plus 1fil%
  &\stablesleft##\stablesright%
  &##\ifstablesright\hfill\fi\internalrule\ifstablesright\else\hfill\fi%
  \tabskip=0pt\cr%
  \ifstablesborderthin%
    \thinline%
  \else%
    \thickline%
  \fi&%
}%
\def\endtable{\advance\stableslines by 1\advance\stablesmode by 1%
   \message{- Rows: \number\stableslines, Columns:  \number\stablesmode>}%
   \stablesel%
   \ifstablesborderthin%
     \thinline%
   \else%
     \thickline%
   \fi%
   \egroup\stablesend%
\endgroup%
\global\stablesinfalse}
%
%

\def\Title#1#2{\rightline{#1}\ifx\answ\bigans\nopagenumbers\pageno0\vskip1in
\else\pageno1\vskip.8in\fi \centerline{\titlefont #2}\vskip .5in}

%
%
\ifx\includegraphics\UnDeFiNeD\message{(NO graphicx.tex, FIGURES WILL BE IGNORED)}
\def\figin#1{\vskip2in}
\else\message{(FIGURES WILL BE INCLUDED)}\def\figin#1{#1}
\fi
\def\Fig#1{Fig.~\the\figno\xdef#1{Fig.~\the\figno}\global\advance\figno
 by1}
%
%
%
%
\def\Ifig#1#2#3#4{
\goodbreak\midinsert
\figin{\centerline{
\includegraphics[width=#4truein]{#3}}}
\narrower\narrower\noindent{\footnotefont
{\bf #1:}  #2\par}
\endinsert
}

\font\ticp=cmcsc10

\def \purge#1 {\textcolor{magenta}{#1}}
\def \new#1 {\textcolor{blue}{#1}}
\def\comment#1{}

\def\\{\cr}
\def\text#1{{\rm #1}}
\def\frac#1#2{{#1\over#2}}

\def\ket#1{| #1\rangle}
\def\bra#1{\langle #1 |}

\def\hf{{1\over 2}} 
\def\O{{\cal O}}

\def\N{{\cal N}}
\def\T{{\cal T}}

\def\I{{\cal I}}
\def\Rw{R\omega}

\def\rs{r_*}
\def\wl{_{\omega l}}
\def\wlm{_{\omega lm}}

\def\vac{|0\rangle}
\def\vacb{\langle0|}
\def\vectr#1{\vec{#1}}
\def\vectl#1{\reflectbox{$\vec{\reflectbox{$#1$}}$}}

\def\calh{{\cal H}}

\def\hnear{{{\cal H}_{\rm near}}}
\def\hfar{{{\cal H}_{\rm far}}}
\def\hbh{{{\cal H}_{\rm bh}}}

\def\roughly#1{\mathrel{\raise.3ex\hbox{$#1$\kern-.75em\lower1ex\hbox{$\sim$}}}}
\font\bbbi=msbm10 
\def\mathbb#1{\hbox{\bbbi #1}}

\def\mthsu{\mathsurround=0pt  }
\def\leftrightarrowfill{$\mthsu \mathord\leftarrow\mkern-6mu\cleaders
  \hbox{$\mkern-2mu \mathord- \mkern-2mu$}\hfill
  \mkern-6mu\mathord\rightarrow$}
\def\overleftrightarrow#1{\vbox{\ialign{##\crcr\leftrightarrowfill\crcr\noalign{\kern-1pt\nointerlineskip}$\hfil\displaystyle{#1}\hfil$\crcr}}}
\overfullrule=0pt

%
%
\lref\AMPS{
  A.~Almheiri, D.~Marolf, J.~Polchinski and J.~Sully,
  ``Black Holes: Complementarity or Firewalls?,''
  JHEP {\bf 1302}, 062 (2013).
  [arXiv:1207.3123 [hep-th]].
}
\lref\AMPSS{
  A.~Almheiri, D.~Marolf, J.~Polchinski, D.~Stanford and J.~Sully,
  ``An Apologia for Firewalls,''
JHEP {\bf 1309}, 018 (2013).
[arXiv:1304.6483 [hep-th]].
}
\lref\BanksHST{
  T.~Banks,
  ``Holographic Space-Time: The Takeaway,''
[arXiv:1109.2435 [hep-th]].
}
\lref\Bousso{
  R.~Bousso,
  ``Firewalls From Double Purity,''
[arXiv:1308.2665 [hep-th]].
}
\lref\Brown{
  A.~R.~Brown,
  ``Tensile Strength and the Mining of Black Holes,''
[arXiv:1207.3342 [gr-qc]].
}
\lref\ChMi{
  P.~L.~Chrzanowski and C.~W.~Misner,
  ``Geodesic synchrotron radiation in the Kerr geometry by the method of asymptotically factorized Green's functions,''
Phys.\ Rev.\ D {\bf 10}, 1701 (1974).
}
\lref\Das{
  S.~R.~Das, G.~W.~Gibbons and S.~D.~Mathur,
  ``Universality of low-energy absorption cross-sections for black holes,''
Phys.\ Rev.\ Lett.\  {\bf 78}, 417 (1997).
[hep-th/9609052].
}
\lref\Deca{
  Y.~Decanini, G.~Esposito-Farese and A.~Folacci,
Phys.\ Rev.\ D {\bf 83}, 044032 (2011).
[arXiv:1101.0781 [gr-qc]].
}
\lref\Dewi{
  B.~S.~DeWitt,
  ``Quantum Field Theory in Curved Space-Time,''
Phys.\ Rept.\  {\bf 19}, 295 (1975).
}
\lref\Mining{
  W.~G.~Unruh and R.~M.~Wald
  ``How to mine energy from a black hole,''
Gen.\ Relat.\ Gravit.\ {\bf 15}, 195 (1983)\semi
  A.~E.~Lawrence and E.~J.~Martinec,
  ``Black hole evaporation along macroscopic strings,''
Phys.\ Rev.\ D {\bf 50}, 2680 (1994)
[hep-th/9312127]\semi
  V.~P.~Frolov and D.~Fursaev,
  ``Mining energy from a black hole by strings,''
Phys.\ Rev.\ D {\bf 63}, 124010 (2001)
[hep-th/0012260]\semi
  V.~P.~Frolov,
 ``Cosmic strings and energy mining from black holes,''
Int.\ J.\ Mod.\ Phys.\ A {\bf 17}, 2673 (2002).
}
\lref\BHMR{
  S.~B.~Giddings,
  ``Black holes and massive remnants,''
Phys.\ Rev.\ D {\bf 46}, 1347 (1992).
[hep-th/9203059].
}
\lref\BHIUN{
  S.~B.~Giddings,
  ``Black hole information, unitarity, and nonlocality,''
Phys.\ Rev.\ D {\bf 74}, 106005 (2006).
[hep-th/0605196].
}
\lref\GiddingsJRA{
  S.~B.~Giddings,
  ``Black holes, quantum information, and the foundations of physics,''
Phys.\ Today {\bf 66}, no. 4, 30 (2013).
}
\lref\BHQIUE{
  S.~B.~Giddings,
  ``Black holes, quantum information, and unitary evolution,''
  Phys.\ Rev.\ D {\bf 85}, 124063 (2012).
[arXiv:1201.1037 [hep-th]].
}
\lref\Models{
  S.~B.~Giddings,
   ``Models for unitary black hole disintegration,''  Phys.\ Rev.\ D {\bf 85}, 044038 (2012)
[arXiv:1108.2015 [hep-th]].
}
\lref\NLvC{
  S.~B.~Giddings,
  ``Nonlocality versus complementarity: A Conservative approach to the information problem,''
Class.\ Quant.\ Grav.\  {\bf 28}, 025002 (2011).
[arXiv:0911.3395 [hep-th]].
}
\lref\NVNLtwo{
  S.~B.~Giddings,
  ``Nonviolent information transfer from black holes: a field theory parameterization,''
Phys.\ Rev.\ D {\bf 88}, 024018 (2013).
[arXiv:1302.2613 [hep-th]].
}
\lref\NVNL{
  S.~B.~Giddings,
  ``Nonviolent nonlocality,''
  Phys.\ Rev.\ D {\bf 88},  064023 (2013).
[arXiv:1211.7070 [hep-th]].
}
\lref\GiddingsPJ{
  S.~B.~Giddings,
  ``Quantum mechanics of black holes,''
  in proceedings of High-energy physics and cosmology, Summer School, Trieste, Italy, June 13-July 29, 1994
[hep-th/9412138].
}
\lref\NLSM{
  S.~B.~Giddings,
  ``Statistical physics of black holes as quantum-mechanical systems,''
[arXiv:1308.3488 [hep-th]].
}
\lref\WABHIP{
  S.~B.~Giddings,
  ``Why aren't black holes infinitely produced?,''
Phys.\ Rev.\ D {\bf 51}, 6860-6869 (1995).
[hep-th/9412159].
}
\lref\SGWIP{
  S.~B.~Giddings,
  work in progress.
}
\lref\GiSh{
  S.~B.~Giddings and Y.~Shi,
  ``Quantum information transfer and models for black hole mechanics,''
Phys.\ Rev.\ D {\bf 87}, 064031 (2013).
[arXiv:1205.4732 [hep-th]].
}
\lref\Hawkunc{
  S.~W.~Hawking,
  ``Breakdown Of Predictability In Gravitational Collapse,''
  Phys.\ Rev.\  D {\bf 14}, 2460 (1976).
}
\lref\Hawk{
  S.~W.~Hawking,
  ``Particle Creation By Black Holes,''
  Commun.\ Math.\ Phys.\  {\bf 43}, 199 (1975)
  [Erratum-ibid.\  {\bf 46}, 206 (1976)].
}
\lref\HaPr{
  P.~Hayden, J.~Preskill,
  ``Black holes as mirrors: Quantum information in random subsystems,''
JHEP {\bf 0709}, 120 (2007).
[arXiv:0708.4025 [hep-th]].
}
\lref\tHooholo{
  G.~'t Hooft,
  ``Dimensional reduction in quantum gravity,''
  [gr-qc/9310026].
}
\lref\tHooftRE{
  G.~'t Hooft,
  ``On the Quantum Structure of a Black Hole,''
Nucl.\ Phys.\ B {\bf 256}, 727 (1985).
}
\lref\MaSu{
  J.~Maldacena and L.~Susskind,
  ``Cool horizons for entangled black holes,''
[arXiv:1306.0533 [hep-th]].
}
\lref\Fuzzref{
  S.~D.~Mathur,
  ``Fuzzballs and the information paradox: A Summary and conjectures,''
[arXiv:0810.4525 [hep-th]].
}
\lref\MathurHF{
  S.~D.~Mathur,
  ``The Information paradox: A Pedagogical introduction,''
Class.\ Quant.\ Grav.\  {\bf 26}, 224001 (2009).
[arXiv:0909.1038 [hep-th]].
}
\lref\PageT{
  D.~N.~Page,
  ``Average entropy of a subsystem,''
Phys.\ Rev.\ Lett.\  {\bf 71}, 1291 (1993).
[gr-qc/9305007]\semi
  D.~N.~Page,
  ``Information in black hole radiation,''
Phys.\ Rev.\ Lett.\  {\bf 71}, 3743 (1993).
[hep-th/9306083].
}
\lref\Pagerev{
  D.~N.~Page,
  ``Black hole information,''
  in 5th Canadian Conference on General Relativity and Relativistic Astrophysics
[hep-th/9305040].
}
\lref\Preskill{
  J.~Preskill,
  ``Do black holes destroy information?,''
  in proceedings of Black holes, membranes, wormholes and superstrings, Houston 1992
[hep-th/9209058].
}
\lref\ARB{
  L.~Susskind,
  ``Black Hole Complementarity and the Harlow-Hayden Conjecture,''
[arXiv:1301.4505 [hep-th]].
}
\lref\Susstrans{
  L.~Susskind,
  ``The Transfer of Entanglement: The Case for Firewalls,''
[arXiv:1210.2098 [hep-th]].
}
\lref\Susstrouble{
  L.~Susskind,
  ``Trouble for remnants,''
[hep-th/9501106].
}
\lref\Sussholo{
  L.~Susskind,
  ``The World as a hologram,''
J.\ Math.\ Phys.\  {\bf 36}, 6377 (1995).
[hep-th/9409089].
}
\lref\Unruh{
  W.~G.~Unruh,
  ``Notes on black hole evaporation,''
Phys.\ Rev.\ D {\bf 14}, 870 (1976).
}
\lref\UnWamine{W.~G.~Unruh and R.~M.~Wald,
  ``Acceleration Radiation and Generalized Second Law of Thermodynamics,''
Phys.\ Rev.\ D {\bf 25}, 942 (1982).
  }
\lref\UnWaunc{
  W.~G.~Unruh and R.~M.~Wald,
  ``On evolution laws taking pure states to mixed states in quantum field theory,''
Phys.\ Rev.\ D {\bf 52}, 2176 (1995).
[hep-th/9503024].
}
\lref\SusskindIF{
  L.~Susskind, L.~Thorlacius and J.~Uglum,
 ``The Stretched horizon and black hole complementarity,''
Phys.\ Rev.\ D {\bf 48}, 3743 (1993).
[hep-th/9306069].
}
\lref\PaRa{
  K.~Papadodimas and S.~Raju,
  ``An Infalling Observer in AdS/CFT,''
[arXiv:1211.6767 [hep-th]].
}
\lref\VerErr{
  E.~Verlinde and H.~Verlinde,
  ``Black Hole Entanglement and Quantum Error Correction,''
[arXiv:1211.6913 [hep-th]].
}
\lref\SGnonpert{
  S.~B.~Giddings,
 ``(Non)perturbative gravity, nonlocality, and nice slices,''
Phys.\ Rev.\ D {\bf 74}, 106009 (2006).
[hep-th/0606146].
}
\lref\BPZ{
  S.~L.~Braunstein, S.~Pirandola and K.~\.Zyczkowski,
  ``Entangled black holes as ciphers of hidden information,''
Physical Review Letters 110, {\bf 101301} (2013).
[arXiv:0907.1190 [quant-ph]].
}
\lref\Star{A.A.~Starobinsky, ``Amplification of waves during reflection from a rotating `black hole'," {\sl Sov. Phys. JETP}, {\bf 37} (1973) 28.}

\Title{
\vbox{\baselineskip12pt  
}}
{\vbox{\centerline{Effective field theory models for nonviolent}\centerline{ information transfer from black holes}
}}

\centerline{{\ticp 
Steven B. Giddings\footnote{$^\ast$}{Email address: giddings@physics.ucsb.edu} and Yinbo Shi\footnote{$^\dagger$}{Email address: yshi@physics.ucsb.edu}
} }
\centerline{\sl Department of Physics}
\centerline{\sl University of California}
\centerline{\sl Santa Barbara, CA 93106}
\vskip.10in
\centerline{\bf Abstract}
Transfer of quantum information from the interior of a black hole to its atmosphere is described, in models based on effective field theory.  This description illustrates that such transfer need not be violent to the semiclassical geometry or to infalling observers, and in particular can avoid producing a singular horizon or ``firewall."  One can specifically quantify the rate of information transfer, and show that a rate necessary to unitarize black hole evaporation produces a relatively mild modification to the stress tensor near the horizon.  In an exterior description of the transfer, the new interactions responsible for it are approximated by ``effective sources" acting on fields in the black hole atmosphere.  If the necessary interactions couple to general modes in the black hole atmosphere, one also finds a straightforward mechanism for information transfer rates to increase when a black hole is mined, avoiding paradoxical behavior.  Correspondence limits are discussed, in the presence of such new interactions, for both small black holes and large ones; the near-horizon description of the latter is approximately that of Rindler space.

\vskip.3in
\Date{}

\newsec{Introduction}

Black hole evaporation\refs{\Hawk} reveals an apparent conflict\foot{For some reviews, see \refs{\Preskill\Pagerev\GiddingsPJ\MathurHF-\GiddingsJRA}.} between the foundational principles of our description of nature via local quantum field theory:  the principles of quantum mechanics, the principles of relativity, and the principle of locality.  Possible resolutions including the abandonment of quantum mechanics\refs{\Hawkunc,\UnWaunc} have been considered, but continued exploration of constraints on consistent scenarios and properties of quantum gravity strongly suggest that locality is a more likely candidate for revision.  Different proposals have been made for modifications to locality, ranging from complementarity/holography\refs{\tHooholo,\Sussholo}, which represents a significant {\it modification to the notion of localization} of information, to the possibility that information escapes a black hole due to new effects with {\it superluminal or nonlocal transfer} of information, 
when described with respect to the semiclassical black hole geometry\refs{\BHMR\BHIUN\NLvC\Models\BHQIUE-\GiSh}.

If the answer is that information leaks out of a black hole due to such new ``nonlocal" effects, this raises a number of questions.  Foremost among them is the question of what more fundamental framework is responsible; spacetime itself may only be emergent from this framework.\foot{For one proposed outline of some features of such dynamics, see \BHQIUE; also see \refs{\BanksHST}.}  Another, more modest, question is how to describe such effects as a correction or modification to the usual semiclassical description of a large black hole.\foot{Though, note that such a description may be no more {\it fundamentally} correct than an attempt to parameterize the evolution of the quantum atom within classical physics.}   Once a black hole has reached a sufficient age, of order its half-life, a very general argument due to Page\refs{\PageT} indicates that the new effects must transfer information at a minimum rate of order one qubit per time $R$, where $R$ denotes the black hole radius.  Such an effect could be comparable in magnitude to the Hawking radiation, which is itself a very small correction to the evolution of a large black hole; this suggests that such modifications are not necessarily implausible.

However, even such ``small" effects have the potential to be dangerous.  It has long been recognized that the Hawking radiation is characterized by the condition that infalling observers crossing the horizon see a near-vacuum state, and this implies specific entanglement between excitations on the two sides of the horizon.  If information is to escape the black hole via some modification of this state that affects the outgoing modes only right at the horizon, then that destroys this entanglement and produces a state that the infalling observer perceives to contain many high-energy particles (this argument was sharpened in \refs{\GiddingsPJ,\BHIUN,\BPZ,\MathurHF,\Models}) or that even destroys the horizon\refs{\BHQIUE}.  Such a picture was advocated as a serious alternative by \refs{\AMPS}, who argue that a sufficiently old but arbitrarily large black hole consequently becomes shrouded in a violent high-energy ``firewall," behind which classical spacetime ceases to exist.

The simplest version of this firewall scenario assumes nonlocal transfer of information:  initially a black hole can form from collapse, but subsequently information transfers from deep within its interior to the horizon, producing the firewall.  In fact, the basic scenario is a limit of the general massive remnant scenario proposed in \BHMR, where the star-like remnant surface that ultimately replaces the horizon lies essentially at the would-be horizon.  The reason for the singular behavior of \AMPS\ is that while such nonlocality is apparently needed, \refs{\AMPS} assumes it stops sharply {\it at} the would-be horizon:  information can nonlocally transfer a distance ten times the radius of the solar system, for the largest known black holes, but not more than a Planck distance further.

Even before \AMPS, this problem was recognized, and a solution was proposed\refs{\Models,\BHQIUE} (see also \BHIUN):  if some effective nonlocality is operative on a scale $\sim R$, then it plausibly allows quantum information to transfer into modes further from the horizon than a Planck distance, and potentially to modes out to a few times $R$, which form the black hole atmosphere.  The transfer need not sharply stop at the stretched horizon.  This suggests a {\it nonviolent} alternative to the firewall proposal advanced in \AMPS.  Specifically, if the information content/entanglement of modes is modified in such a soft, long-distance fashion, this does not necessarily produce particles that the infalling observer sees as damaging, or that destroy the horizon.    The basic underlying assumption of this scenario is thus that the unitarity-restoring corrections preserve the classical picture of the near-horizon spacetime, to a good approximation, but may modify the outgoing radiation, in order to transfer information, in a manner that does not do violence to this picture.  This is specifically a violation of axiom 2 of black hole complementarity\SusskindIF, stating that evolution outside the horizon is described by local quantum field theory. This scenario is less radical than that of \AMPS\ both in being nonviolent, and in not requiring fine-tuning of the nonlocal transfer.  The latter is plausible, particularly given that we may not know precisely where the horizon is; instead the nonlocal information transfer ranges over a characteristic scale $\sim R$.

To be believed, such a scenario needs to be subject to some consistency tests.    The problem of describing restoration of unitarity is remarkably constrained -- so much so that, as we have outlined, certain assumptions lead to unphysical behavior\AMPS.  An important -- and sharp -- question is thus whether there is  ``room" for consistent modification of local quantum field theory 
that describes the quantum information transfer necessary to save quantum mechanics, while at the same time also preserving an approximate semiclassical picture.

A first step regarding such tests was giving more detailed models for the proposed behavior\refs{\NVNL,\NVNLtwo}.  Ref.~\NVNLtwo\ in particular suggested modeling the physics in an effective field theory framework, but with additional interactions that accomplish the transfer of quantum information needed to save unitary evolution.  Such a model gives a way to check various possible features of such a scenario.  One aspect to be checked is that of nonviolence -- if the new interactions are sufficiently large to transfer the needed information, for example at the minimum rate described above, we would like to verify that they do not lead to large effects unduly damaging infalling observers or the horizon.  One would also like to check that such a picture also gives a non-problematic story in the presence of black hole mining\refs{\UnWamine, \Mining}, which provides an important test by enhancing black hole decay rates.  Another question regards {\it correspondence}:  in the large-$R$ limit, where the vicinity of a black-hole horizon approaches flat space, one expects the description of observations of stationary observers to match onto the usual field-theory description of accelerated observers\refs{\Unruh}.   Yet another set of constraints come from the need for a consistent statistical/thermodynamic description\refs{\NVNL,\AMPSS,\NLSM}, where one in particular finds that generic enhancement of the black hole disintegration rate due to the extra interactions indicates a black hole entropy smaller than that given by Bekenstein and Hawking.  

Responses to the first two questions -- regarding nonviolence and mining -- were outlined in \NVNLtwo, and will be provided in further detail here.  Specifically, after giving a more detailed description of models for the proposed interactions and of black hole metrics and modes, section two demonstrates the effect of a simple example of such interactions on fields surrounding a black hole.  Section three then investigates the asymptotics of the resulting excitations, and the resulting stress tensor, both at null infinity, and in the vicinity of the horizon.  The latter shows that for a wide class of interactions, the effect near the horizon is indeed nonviolent.  Specifically, section four shows that if the asymptotic flux of excitations is the benchmark size to unitarize black hole disintegration, there is a corresponding modest increase in the energy density in modes near the horizon.  This energy density decreases with increasing $R$ -- providing a test of correspondence.  

Moreover, the new interactions are generically expected to couple to modes with various angular momenta.  If they do so with roughly uniform strength for higher partial waves, there is very little effect on the black hole decay rate, due to large gray-body suppression factors for asymptotic radiation.  But, if mining apparata are introduced into the black hole atmosphere, providing an additional channel for excitations to escape, there is a commensurate increase in the rate that the interactions can transfer information to outgoing modes\NVNLtwo.  Further details of this important consistency check in the presence of mining -- which demonstrates a natural mechanism to avoid the potential problem of ``overfilling" black holes with information -- are also provided in section four.  Section five closes with discussion of generalizations of the simplified models explicitly treated in this paper and with brief discussion of the generic extra energy flux, and then returns to elaborate on the important question of correspondence.  An appendix gives a WKB estimate of relevant black hole gray-body factors.

\newsec{New interactions and their effects}

\subsec{The effective-source approximation}

It has seemed increasingly apparent that local quantum field theory (LQFT) cannot give a unitary description of black hole evolution, and that we must seek a different, and more fundamental, framework.  If that framework respects the principles of quantum mechanics, one promising approach to its formulation is through a structure of nested and overlapping quantum subsystems, giving a version of localization that might approximate that of LQFT\refs{\BHQIUE}.  For example, the Hilbert space describing a black hole and its environment might be contained in a product of the form\refs{\Models,\BHQIUE}
\eqn\prodstruct{\calh\subset \hbh\otimes\hnear\otimes\hfar\ ,}
where we have separate subsystems for the black hole, the near black hole ``atmosphere," and states asymptotically far from the black hole.  Further refinement of the subsystem structure is also expected to be possible (see {\it e.g.} \refs{\NLSM}).  
For a big black hole and for many purposes, the states of this Hilbert space and evolution should be well-approximated by LQFT.

Of course, a departure from LQFT that apparently must become important for even a large black hole is transfer of information\refs{\Models\BHQIUE-\GiSh} from the internal states of the black hole to degrees of freedom that escape to infinity.  For a sufficiently old black hole, of radius $R$, such transfer must take place at a minimum rate of at least one qubit per time $R$.  
Such transfer can be described in terms of unitary evolution with an infinitesimal generator including terms of the form\NVNL\
\eqn\waveH{ H_{trans} \sim {1\over R} a^\dagger_{near}{\cal N} a_{bh} + h.c.\ , }
with operators acting to annihilate excitations in $\hbh$ and create those in $\hnear$, or vice versa ($\N$ is a transfer matrix).
Alternatively, such dynamics could be described by introducing bilocal\foot{Higher-order terms may also be present.} contributions to the action\NVNLtwo,
\eqn\genact{S_{NL}= \sum_{AB} {\cal O}_A G_{AB}  {\cal O}_B\ ,}
where ${\cal O}_A$ are operators acting on $\hbh$, ${\cal O}_B$ are operators acting on $\hnear$,  and $G_{AB}$ are coefficients describing the propagation between the two.\foot{A possible straightforward generalization is transfer to $\hfar$, but this involves a more significant departure from usual locality and will not be developed in this paper.  Note in particular that there are many more low-energy modes available at long distance that could carry the information, and that these could be {\it e.g.} populated at low temperature.  These are not ordinarily accessed near the black hole, due to the centrifugal barrier.  But, nonlocal transfer to scales $\gg R$ -- if present -- would avoid this restriction.  Also, $\O_B$ in \genact\ may be generalized to act on ``degrees of freedom" just inside the horizon, in a more refined description\refs{\NLSM}.}  

For a big black hole over sufficiently short times, we expect that the states $\hnear$ of the atmosphere can be well-approximated via LQFT, and in particular that the operators in \genact\ acting on $\hnear$ can be replaced by local operators of the theory, ${\O}_B\rightarrow {\cal O}_b(x)$. 
While terms like \waveH\ or \genact\ 
need to give 
an $\O(1)$ perturbation to the Hawking process, the latter is a very small effect for a large black hole.  This suggests that interactions of the required size can be treated as a perturbative correction to the description of the dynamics via LQFT in a semiclassical background\NVNLtwo. This evolution is in particular nonlocal with respect to the causal structure defined by the semiclassical background geometry.

While understanding the full unitary quantum dynamics is clearly very important, there are also important questions that largely depend only on how the dynamics act on states near the horizon.  In particular, there has been longstanding awareness, sharpened in \refs{\GiddingsPJ, \BHIUN,\MathurHF,\Models,\BHQIUE}, that interactions that transfer information from the black hole interior to short-wavelength excitations near the horizon produce high-energy particles as seen by the infalling observer, and are typically expected to destroy the horizon.  To avoid such violence, \refs{\Models,\BHQIUE,\NVNL} postulated that the information transfer (which can be characterized in terms of entanglement transfer\refs{\HaPr,\GiSh,\Susstrans}) is instead to excitations at longer wavelengths, up to scales\foot{More generally the relevant wavelengths could be of size $R^p$, with $0<p\leq1$.} 
$\sim R$.  

The question of whether  nonviolent information transfer to such longer-wavelength modes can be accomplished, with sufficient magnitude to restore unitarity to black hole disintegration, and without destroying the horizon or infalling observers, is largely dependent on how interactions such as \genact\ act on the state outside the black hole.  For the purposes of investigating this question, one may make an additional approximation, and replace the operators in \genact\ that depend on the internal state of the black hole by sources in the external field-theory action:
\eqn\effsource{S_{NL}\rightarrow\sum_{Ab}\int dV_4 {\cal O}_A G_{Ab}(x)  {\cal O}_b(x)\rightarrow \sum_b\int dV_4 J_b(x){\cal O}_b(x)\ ,}
where $dV_4$ is the volume element and $\O_b(x)$ acts on fields near the black hole.  While in the more fundamental description \genact\ the sources $J_b$ correspond to operators dependent on the black hole internal state and dynamics, for investigating the information-relaying capacity of such interactions, and characterizing their effects on modes and observers near a black hole horizon, these sources may for many purposes be approximated as external, classical sources.  We refer to this as the  {\it effective source} approximation.

Ultimately the unitary mechanics underlying quantum gravity should determine the interactions \genact\ and which operators they couple to in an effective description \effsource.  Given the universality of gravity -- and the need to conserve gauge charges -- one interesting possibility is a coupling of the form $J^{\mu\nu}T_{\mu\nu}$. 
However, to investigate basic features of such interactions, for present purposes we consider linear couplings to field operators. As we will find, such couplings illustrate important points of principle, and in particular the possibility of transmitting the necessary information without doing violence to the horizon or to infalling observers.

For simplicity, let us consider  a single massless scalar field with action
\eqn\phiact{S_\phi = -\hf\int dV_4 \left(\nabla \phi \right)^2\ .}
 In this paper we will consider the simple model of an effective source that couples linearly to this scalar field, through a term in the lagrangian
\eqn\Jcoup{S_J = -\int dV_4 J(x) \phi(x)\ .}
Important questions will include 1) what $J(x)$ would produce sufficient excitation to carry out the quantum information necessary to unitarize black hole disintegration, including in the possible presence of black hole mining\refs{\UnWamine,\Mining,\AMPS}, and 2) what effects does such a $J(x)$ have on the atmosphere of the black hole, and on observers falling through that atmosphere.  

\subsec{Stress tensor from $J$}

A first approach to answering the preceding questions is to find the quantum stress tensor resulting from a source like \Jcoup.  The  stress tensor for the scalar field $\phi$ takes the form
\eqn\Tdef{T_{\mu\nu} = -\frac{2}{\sqrt{-g}}\frac{\delta S[\phi]}{\delta g^{\mu\nu}} = \partial_\mu \phi \partial_\nu \phi - \frac{g_{\mu\nu}}{2} \left[ \left(\partial \phi \right)^2 + 2J\phi \right]\ .}
Before the source \Jcoup\ is introduced, we assume that the black hole is in a state $\vac$ which could be either the Unruh or Hartle-Hawking vacuum.  Such a vacuum results in an outgoing Hawking flux, which can be seen by calculating, with a  careful regulator,
\eqn\Tvev{\vacb T_{\mu\nu} \vac = \T_{\mu\nu}\ .}
The effect of the source \Jcoup\ can be described by treating it as a perturbation, and working in the interaction picture.  In its presence, the state outside the black hole becomes
\eqn\Jstate{|J,t\rangle = T\exp\Big\{-i\int^t dV'_4 J(x') \phi(x')\Big\}\vac\ ,}
where time ordering is performed with respect to a choice of time slicing of the exterior geometry of the black hole.  
For such a linear coupling in the field, the time ordering can be removed at the price of a c-number phase $\beta(t)$ (see appendix):
\eqn\Jket{|J,t\rangle = e^{i\beta(t)} \exp\Big\{-i\int^t dV'_4 J(x') \phi(x')\Big\}\vac\ .}
For both the Unruh and Hartle-Hawking vacua, the field has vanishing expectation value, $\langle0|\phi(x)|0\rangle=0$.  However, with the source the field picks up an expectation value,
\eqn\defphiJ{ \eqalign{
\phi_J(x) &\equiv \bra{J,t}\phi(x)\ket{J,t} \\
&= \bra{0}\phi(x)\ket{0} + \bra{0} \left[\phi(x), -i\int^t dV'_4 J(x')\phi(x')\right] \ket{0} \\
&= \int dV'_4  G_R(x,x')J(x')\ ,
}}
where the retarded Green function is
\eqn\defgreen{ G_R(x,x') \equiv -i\theta(t-t') \left[\phi(x), \phi(x')\right] \ .}
Note that $\phi_J$ behaves like a classical field; in particular, due to vanishing equal-time commutators, $\partial_\mu\phi_J(x)$ is equal to $\bra{J,t}\partial_\mu\phi(x)\ket{J,t}$.  The two-point functions in \Tdef\ then have a simple form, following from
\eqn\twoptfunc{ \eqalign{
&e^{i\int^t J \phi} \partial_\mu \phi(x) \partial_\nu \phi(x) e^{-i\int^t J \phi}  \\
&= \left[ e^{i\int^t J \phi}  \partial_\mu \phi(x) e^{-i\int^t J \phi} \right]
	\left[ e^{i\int^t J \phi}  \partial_\nu \phi(x)  e^{-i\int^t J \phi} \right]  \\
&= \left[ \partial_\mu\phi(x) + \partial_\mu\phi_J(x) \right] \left[ \partial_\nu\phi(x) + \partial_\nu\phi_J(x) \right] \ . \\
}}
The change of the expectation value of the stress tensor \Tdef\ due to $J$ then follows
\eqn\TinJ{ \eqalign{
\bra{J,t} T_{\mu\nu} \ket{J,t} &= \bra{0} e^{i\int^t J \phi}
	\left[ \partial_\mu\phi\partial_\nu\phi - \frac{1}{2}g_{\mu\nu} \left(g^{\rho\sigma}\partial_\rho\phi\partial_\sigma\phi +2J\phi\right)\right]
	e^{-i\int^t J \phi}\ket{0} \\
&= \T_{\mu\nu} + T_{\mu\nu}[\phi_J]\ , \\
}}
where $T_{\mu\nu}[\phi_J]$ is \Tdef\ evaluated with $\phi=\phi_J$ given by \defphiJ.  This gives the extra flux resulting from $J$, which is similar to that of a classical field on top of a quantum background.

Equation \TinJ\ has an important implication.  Specifically, such a classical field produces a {\it positive} flux of energy at infinity.  This means that extra interactions like \Jcoup\ would increase the decay rate of the black hole above the Hawking rate\refs{\BHQIUE,\GiSh,\NVNL,\NVNLtwo}.  Such an extra flux has 
 potentially important consequences for black hole statistical mechanics\refs{\NLSM}.

\subsec{Black hole metric and modes}

In order to describe the properties of the state \Jstate\ and its energy-momentum \TinJ\ in a very explicit example, we consider the Schwarzschild geometry, and modes propagating on this background.  The metric is
\eqn\Schw{ds^2 = -f(r)dt^2 + {dr^2\over f(r)} + r^2 d\Omega^2\ .}
Specifically, considering a four-dimensional black hole\foot{Most of our results are readily generalized to higher-dimensional Schwarzschild.} with Schwarzschild radius $R$,
\eqn\f{f = 1 - \frac{R}{r}\ .}

Modes propagating in this background are simply understood by introducing  tortoise coordinates, in which the metric takes the form
\eqn\metric{ds^2 = f(\rs)(-dt^2 + d\rs^2) + r^2(\rs)d\Omega^2\ .}
The tortoise coordinate is thus defined by
\eqn\deftort{r_*=\int {dr\over f(r)}\ .}
There is an arbitrary integration constant, chosen for later simplicity; this choice differs slightly from the traditional one, and specifically is defined via
\eqn\defr{ \eqalign{
	e^{\rs/R} &= \left(\frac{r}{R} -1\right) e^{r/R -1}\\
	\frac{r}{R} -1 & = W\left(e^{\rs/R}\right) \ ,
}}
where $W$ is Lambert's W function.\foot{  $W(z)$ is the principal branch of $z = W(z)e^{W(z)}$.}
For later convenience, we may also introduce null coordinates $x^\pm = t \pm \rs$, in which the metric is
\eqn\metrictwo{ds^2 = -f(\rs)dx^+dx^- + r^2(\rs)d\Omega^2\ .}
%

Solutions to the equation of motion $\nabla^2 \phi =0$ for a free scalar field in the coordinates \metric\ can be expanded in a mode expansion of the form
\eqn\modeexp{
	\phi(x) = \sum_{Alm}\int_0^\infty \frac{d\omega}{2\pi2\omega} \left[U^A_{\omega lm}(x) b^A_{\omega lm} + \text{h.c.} \right]\ , }
\eqn\modes{	
	U^A_{\omega lm} = u^A_{\omega l}(r_*) e^{-i\omega t} \frac{Y_{lm}(\Omega)}{r} \ .}
In this expansion, $b^A_{\omega lm}$ are arbitrary coefficients, and the radial wavefunctions $u^A_{\omega l}(r_*)$ arise from solutions of a $1+1$-dimensional flat-space wave equation in $r_*$ and $t$, with an effective potential, 
\eqn\defu{ \left(\frac{\partial^2}{\partial\rs^2}+\omega^2\right) u^A\wl = V_l u^A\wl \ ,}
\eqn\defV{ V_l = f(\rs) \left[ \frac{l(l+1)}{r^2} + \frac{R}{r^3} \right]  \ .}

\Ifig{\Fig\AsympModeDiag}{Schematic of the different bases for modes.  The black curve represents the potential.  {\it Past} modes are purely incoming from $r_*=\pm\infty$ in the asymptotic past; in the future, they have both reflected and transmitted parts from the potential barrier.  {\it Future} modes are likewise purely outgoing to $r_*=\pm\infty$ in the asymptotic future.  The past and future bases are related by complex conjugation.}{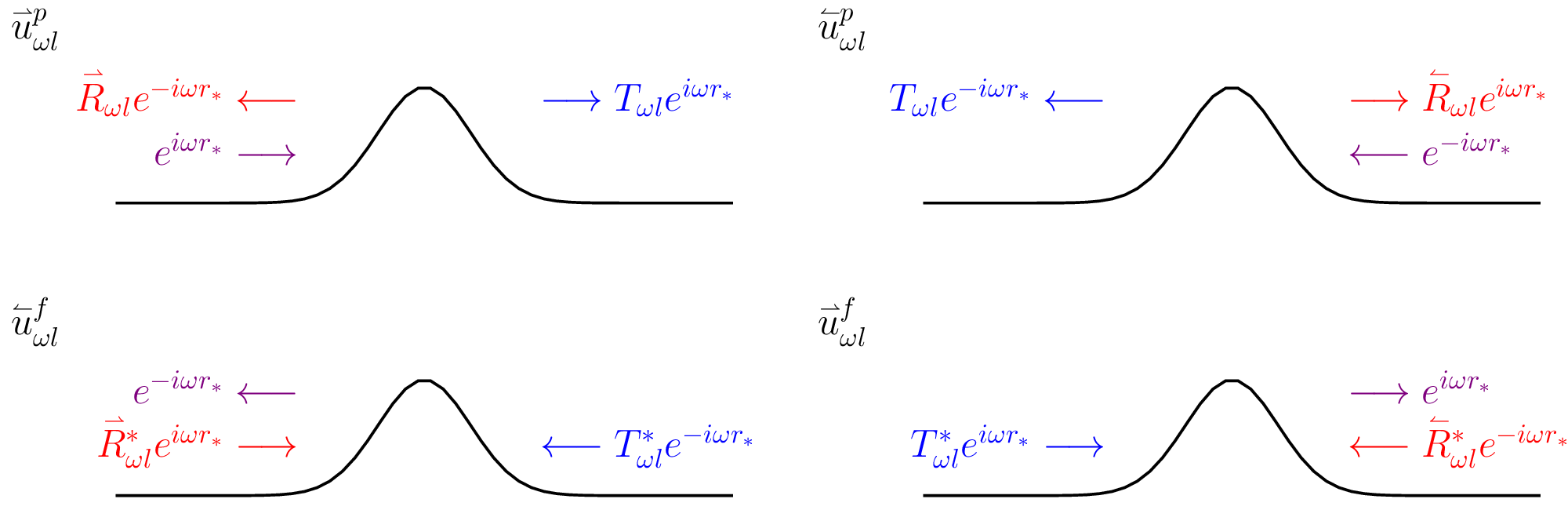}{6}

Different bases for solutions of \defu, labeled by the index $A$, may be chosen\refs{\ChMi,\Dewi}, as illustrated in \AsympModeDiag.  One basis is the {\it past modes} (with simple behavior in the asymptotic past), for which $A \in (p\rightarrow,p\leftarrow)$, and another basis is the {\it future modes} (with simple behavior in the asymptotic future), 
with $A \in (f\rightarrow,f\leftarrow)$.  Specifically, these bases have asymptotic behavior (with names as in \refs{\ChMi})
%
%
\eqn\uasymp{ \begintable
\vt $r_* \rightarrow -\infty $ \vt $ r_* \rightarrow \infty $ \elt
$ \vectr{u}^p \text{ (up)} $ \vt $ e^{i\omega r_*} + \vectr{R}_{\omega l}e^{-i\omega r_*} $ \vt $ T_{\omega l}e^{i\omega r_*} $ \elt
$ \vectl{u}^p \text{ (in)} $ \vt $ T_{\omega l}e^{-i\omega r_*}$ \vt $ e^{-i\omega r_*} + \vectl{R}_{\omega l}e^{i\omega r_*} $ \elt
$ \vectl{u}^f =\vectr{u}^{*p} \text{ (down)} $ \vt $ e^{-i\omega r_*} + \vectr{R}^*_{\omega l}e^{i\omega r_*} $ \vt $ T^*_{\omega l}e^{-i\omega r_*} $ \elt
$  \vectr{u}^f=\vectl{u}^{* p}\text{ (out)} $ \vt $ T^*_{\omega l}e^{i\omega r_*}$ \vt $ e^{i\omega r_*} + \vectl{R}^{*}_{\omega l}e^{-i\omega r_*}$
\endtable
}
Different bases are useful depending on the physical question being asked.

Quantization of $\phi$ is performed with the following conventions.  The modes \modes\ have been chosen to have Klein-Gordon norm 
\eqn\KG{ \eqalign{
(U^A_{\omega lm}, U^{A'}_{\omega' l'm'}) = i \int r^2dr_* d\Omega U^{A*}_{\omega lm}{\overleftrightarrow \partial}_t U^{A'}_{\omega' l'm'} = 
 2\pi2\omega\delta(\omega-\omega')\delta_{ll'} \delta_{mm'} \delta_{AA'} \ ,
}}
as seen {\it e.g.} from the asymptotic behavior in \uasymp, where $A,A'$ are chosen to range over either past modes, or over future modes.  The canonical commutation relations are
\eqn\CCR{[\partial_t \phi(x),\phi(x')] = -i\delta(r_*-r_*') {\delta^2(\Omega-\Omega')\over r^2}\ ,}
and result in commutators
\eqn\comm{ [b^A\wlm, b_{\omega' l'm'}^{A'\dagger}] = 2\pi2\omega\delta(\omega-\omega')\delta_{ll'} \delta_{mm'} \delta_{AA'} \ .}

\subsec{Calculating $\phi_J$}

We next calculate $\phi_J$, using these mode expansions; this in turn determines the stress tensor, through \TinJ.

Specifically, from the mode expansion \modeexp\ and the commutators \comm, eq.~\defgreen\ determines the retarded Green function as
\eqn\greentwo{ G_R(x,x') = -i\theta(t-t') \sum_{Alm}\int \frac{d\omega}{2\pi2\omega} \left[ U^A_{\omega lm}(x) U^{A*}_{\omega lm}(x') - \text{c.c.} \right] \ .}
(Unless otherwise noted, $\omega$ integrals are over the positive reals, and all other integrals are over the full domain  -- {\it e.g.} reals for one-dimensional  integrals or $\mathbb{R}^4$ for volume integrals.)
%
Thus, from \defphiJ, $\phi_J$ becomes
\eqn\calcphiJ{ \eqalign{
\phi_J(x) &= -i\int^t dV'_4 J(x') \sum_{Alm} \int\frac{d\omega}{2\pi2\omega}
	\left[ U^A_{\omega lm}(x) U^{A*}_{\omega lm}(x') - \text{c.c.} \right] \\
&= -i \sum_{Alm} \int\frac{d\omega}{2\pi2\omega}
	\left[\alpha^A_{\omega lm}(t) U^A_{\omega lm}(x)  - \text{c.c.} \right] \ ,
}}
with coefficients $\alpha$ defined as
\eqn\defalpha{ \alpha^A_{\omega lm}(t) =  \int^t dV'_4 U^{A*}_{\omega lm}(x')J(x') \ . }
Let $J$ be given by the mode expansion
\eqn\jexp{ J(x) = \sum_{lm} \int \frac{d\omega}{2\pi} j_{\omega lm}(r) e^{-i\omega t} \frac{Y_{lm}(\Omega)}{r} + \text{c.c.} \ , }
and introduce the notation  
\eqn\defbracket{ \langle a(r),b(r)\rangle = \int_{-\infty}^{\infty} f a^*(r)b(r) dr_* = \int^\infty_R dr a^*(r)b(r) \ .}
Then, the coefficients become
\eqn\calcalpha{ \eqalign{
\alpha^A_{\omega lm}(t) 
&= \int^t dt' \int \frac{d\omega'}{2\pi} \left[
	\langle u^A_{\omega l},j_{\omega' lm}\rangle  e^{i(\omega-\omega') t'} + (-1)^m\langle u^A_{\omega l},j^*_{\omega' l-m}\rangle  e^{i(\omega+\omega') t'}  \right] \\
&= \int \frac{d\omega'}{2\pi} \left[ \langle u^A_{\omega l},j_{\omega' lm}\rangle  \frac{ e^{i(\omega-\omega') t}}{i(\omega-\omega')+\epsilon} + (-1)^m \langle u^A_{\omega l},j^*_{\omega' l-m}\rangle  \frac{e^{i(\omega+\omega') t}}{i(\omega+\omega')+\epsilon} \right] \ ,
}}
where in the last equality we introduce the small convergence factor $\epsilon>0$ to  regulate the integrals.
Thus, the expression \calcphiJ\ for $\phi_J$ becomes
\eqn\calcphiJ{ \eqalign{
\phi_J(x) 
&= - \sum_{Alm} \int \frac{d\omega}{2\pi2\omega}\frac{d\omega'}{2\pi}    \left[ {\langle u^A_{\omega l},j_{\omega' lm}\rangle \over \omega-\omega'-i\epsilon}  u^A_{\omega l}(r)  e^{-i\omega' t}{Y_{lm}(\Omega)\over r}  \right.   \\
	& \qquad +  \left. (-1)^m {\langle u^A_{\omega l},j^*_{\omega' l-m}\rangle\over \omega+\omega'-i\epsilon}  u^A_{\omega l}(r) e^{i\omega' t}{Y_{lm}(\Omega)\over r} + \text{c.c.}  \right] \ .
}}

\newsec{Asymptotics}

\subsec{$\phi_J$}

We would next like to determine the asymptotic form of $\phi_J$, and the corresponding stress tensor, both at $r,r_*\rightarrow\infty$ and near the horizon, $r_*\rightarrow -\infty$.  
First consider $r_*\rightarrow\infty$.  The asymptotic form can be found by using the future basis.  Inserting its asymptotic behavior \uasymp\ into \calcphiJ\ and using the coordinates $x^\pm$ of \metrictwo\ gives
\eqn\lk{ \eqalign{
\phi_J &\rightarrow - \sum_{lm} \int \frac{d\omega}{2\pi2\omega }\frac{d\omega'}{2\pi} \left\{ {Y_{lm} \over r}
 \left[\left(
	\langle \vectl{u}^f_{\omega l},j_{\omega' lm}\rangle T^*_{\omega l} +\langle \vectr{u}^f_{\omega l},j_{\omega' lm}\rangle \vectl{R}^*_{\omega l}\right)\frac{e^{i(\omega - \omega')(-r_*)}  e^{-i\omega' x^+}}{\omega-\omega'-i\epsilon} \right. \right.  \\
	& \qquad + (-1)^m \left( \langle \vectl{u}^f_{\omega l},j^*_{\omega' l-m}\rangle T^*_{\omega l}+\langle \vectr{u}^f_{\omega l},j^*_{\omega' l-m}\rangle \vectl{R}^*_{\omega l}\right) \frac{e^{i(\omega + \omega')(-r_*)} e^{i\omega' x^+}}{\omega+\omega'-i\epsilon}  \\
	& \qquad + \langle \vectr{u}^f_{\omega l},j_{\omega' lm}\rangle  \frac{e^{i(\omega - \omega')r_*}  e^{-i\omega' x^-}}{\omega-\omega'-i\epsilon}  
	+ \left. \left. (-1)^m \langle \vectr{u}^f_{\omega l},j^*_{\omega' l-m}\rangle \frac{e^{i(\omega + \omega')r_*} e^{i\omega' x^-}}{\omega+\omega'-i\epsilon} \right] + \text{c.c.} \right\}\ .}}
This expression is simplified using the distributional identities:
\eqn\k{ \eqalign{
2\pi \delta(\omega) &= \lim_{t \rightarrow \infty}  \frac{-i e^{i\omega t}}{\omega-i\epsilon} \\
0 &= \lim_{t \rightarrow -\infty} \frac{-i e^{i\omega t}}{\omega-i\epsilon}\ .
}}
The second of these implies vanishing of the first and second rows of \lk, and the first, together with $\omega,\omega'>0$, implies vanishing of the last term of \lk, giving  the $r_*\rightarrow\infty$ result
%
%
\eqn\lkt{\phi_J\rightarrow -i \sum_{lm} \int \frac{d\omega}{2\pi2\omega} \left[ {Y_{lm}\over r} \langle \vectr{u}^f_{\omega l},j_{\omega lm}\rangle e^{-i\omega x^-} - \text{c.c.} \right]\ .}
%
%
%
%
%
%
%
%
Similar steps can be applied to derive the behavior as $r_*\rightarrow-\infty$:
\eqn\lkjjk{ 
\phi_J\rightarrow -i \sum_{lm} \int \frac{d\omega}{2\pi2\omega } \left[ {Y_{lm}\over R} \langle \vectl{u}^f_{\omega l},j_{\omega lm}\rangle  e^{-i\omega x^+} - \text{c.c.} \right]\ .
}

\subsec{Stress tensor}

Let us first consider the asymptotic form of the stress tensor $T[\phi_J]$ at infinity, $r_*\rightarrow\infty$.  Specifically, the outgoing flux is given by the components $T_{--}$,  in the coordinates $x^\pm$.  The integrated flux follows from \TinJ\ and \lkt,  
\eqn\Tmm{ \eqalign{
\int dtT_{--}  & \rightarrow \int dt \left( \sum_{lm} \int \frac{d\omega}{4\pi} \left[ \frac{Y_{lm}}{r} \langle \vectr{u}^f_{\omega l},j_{\omega lm}\rangle  e^{-i\omega x^-} + cc \right] \right)^2 \\
&= \sum_{ll'mm'} \frac{Y_{lm}Y^*_{l'm'}}{r^2} \int \frac{d\omega}{4\pi} \langle \vectr{u}^f_{\omega l},j_{\omega lm}\rangle
	\langle \vectr{u}^f_{\omega l'},j_{\omega l'm'}\rangle^*\ ,
}}
and 
integrating over angles yields the total radiated energy
\eqn\Erad{
E = \int_{r\gg R} dt\, r^2d\Omega T_{--}  = \sum_{lm} \int \frac{d\omega}{4\pi} |\langle \vectr{u}^f_{\omega l},j_{\omega lm}\rangle|^2\ .
}

The source $J$ also produces a flux into the black hole, which may be found by similarly computing the $r_*\rightarrow-\infty$ behavior of $T_{++}$, using \lkjjk.  This gives integrated flux
\eqn\Tpp{ \eqalign{
\int dt T_{++}  & \rightarrow \int dt \left( \sum_{lm} \int \frac{d\omega}{4\pi} \left[ \frac{Y_{lm}}{R} \langle \vectl{u}^f_{\omega l},j_{\omega lm}\rangle  e^{-i\omega x^+} + cc \right] \right)^2 \\
&= \sum_{ll'mm'} \frac{Y_{lm}Y^*_{l'm'}}{R^2} \int \frac{d\omega}{4\pi} \langle \vectl{u}^f_{\omega l},j_{\omega lm}\rangle
	\langle \vectl{u}^f_{\omega l'},j_{\omega l'm'}\rangle^*\ ,
}}
and total absorbed energy 
\eqn\Eabs{
E = \int_{r=R} dt\, R^2 d\Omega T_{++} = \sum_{lm} \int \frac{d\omega}{4\pi} |\langle \vectl{u}^f_{\omega l},j_{\omega lm}\rangle|^2\ .
}

We will investigate the size of these fluxes in the next section, in scenarios where the outward flux is large enough to carry the needed information away from the black hole.  But, before doing that, there is another important check.  Specifically, if there is an outward flux present that is traceable back to the horizon, due to infinite blueshift that flux becomes singular at the horizon, as described in \refs{\GiddingsPJ, \BHIUN, \MathurHF, \Models, \BHQIUE, \AMPS}.  Thus, to parameterize a ``nonviolent" scenario where the horizon is regular, as seen by infalling observers, we need to check that the $J$'s we consider do not produce such a singular flux.

\subsec{Nonviolent horizon}

The infinite blueshift witnessed by infalling observers is readily understood by transforming from the $x^\pm$ coordinates to Kruskal coordinates $X^\pm$, through the relation
\eqn\defkrusk{
X^\pm = \pm 2R e^{\frac{\pm x^\pm}{2R} \ .}
}
While the $x^-$ coordinates are singular at the future horizon, the Kruskal coordinates are non-singular coordinates for observers falling through the horizon.  From \defkrusk, we find $\partial X^-/\partial x^- = e^{-x^-/2R} = -X^-/2R$.  Thus,
\eqn\Tkruskt{T^{\rm Krusk}_{--} = \left({2R\over X^-}\right)^2 T_{--}}
will be singular unless the outward flux $T_{--}$ vanishes at least as rapidly as $(X^-)^2$ at the horizon, $X^-=0$.  

To check this, we  examine the behavior of 
\eqn\phikrusk{\partial_{X^-} \phi_J = e^{x^-/2R} \partial_- \phi_J\ }
 near the horizon.  $\phi_J$ satisfies the classical equation of motion,
\eqn\eomphiJ{
\nabla^2 \phi_J = J\ .}
Expanding in partial waves,
\eqn\PWexp{\phi_J = \sum_{lm} \phi_{lm}(t,r_*) \frac{Y_{lm}(\Omega)}{r} \quad ,\quad J=\sum_{lm} j_{lm} \frac{Y_{lm}(\Omega)}{r}\ ,}
this becomes
\eqn\PWeqn{\left[ -\partial_t^2 + \partial_{r_*}^2 - V_l(r) \right] \phi_{lm} = f(r) j_{lm}\ ,}
with $f$ given in \f\ and $V_l$ given in \defV.  This reduces the problem to a collection of 1+1-dimensional problems.  To reduce clutter, we will fix $l, m$ for the remainder of this section, and suppress these subscripts.  Thus \PWeqn\ becomes
\eqn\PWsimp{ -4\partial_+\partial_- \phi -V\phi = fj\ .}
With the problem rewritten in terms of the potential \PWeqn, \PWsimp, the basic idea is that at a fixed point $(t,r_{*0})$ near the horizon, the right-moving piece $\partial_-\phi$ receives contributions from two places:  the source $J$ that is located to the left of $r_{*0}$, and left-moving waves coming from the right of $r_{*0}$ that then reflect off of the potential $V$ and become right-moving.  Since the potential is small near the horizon (see \defV), we will treat it perturbatively, and correspondingly expand $\phi =\Phi_0+\Phi_1+\cdots$.

To zeroth order in $V$, \PWsimp\ has solution 
\eqn\zeroorderjr{ 
\partial_- \Phi_0 = -\frac{1}{4} \int_{-\infty}^{x^+} dx^+ f j =-\frac{1}{4}e^{-x^-\over 2R}  \int_{-\infty}^{x^+}dx^+\frac{R}{r} e^{1-r/R} e^{x^+\over 2R} j\ ,}
where we have used \defr.  This implies
\eqn\Jfinite{
\partial_{X^-} \Phi_0(x^-,x^+) = -\frac{1}{4}  \int_{-\infty}^{x^+}dx^+\frac{R}{r} e^{1-r/R} e^{x^+\over 2R} j(x^-,x^+)}
is finite, {\it i.e.} the horizon is regular, as long as the  latter integral is finite, which will be true if $J(x^-,x^+)$ is smaller than  $\exp\{-x^+/(2R)\}$ as $x^+\rightarrow -\infty$.

The first-order equation is
\eqn\firstorder{-4\partial_+\partial_- \Phi_1= V \Phi_0\ ,}
which likewise implies 
\eqn\FOsoln{
\partial_{X^-}\Phi_1 = -\frac{1}{4}  \int_{-\infty}^{x^+}dx^+\frac{R}{r} e^{1-r/R} e^{x^+\over 2R} {V\over f}\Phi_0\ .}
In this equation, the $r$-dependent factors are approximately finite constants near the horizon (see \defV), and the integral converges for any finite $\Phi_0$.  One may likewise proceed to find finite higher-order contributions to the solution.\foot{These include contributions of size comparable to \FOsoln, due to reflection from $V$ at $r>r_0$.}  We see from \PWexp\ that $\partial_{X^-}\phi_J$ has an additional term,
\eqn\philim{\partial_{X^-} \phi_J = \sum_{lm} \left( \partial_{X^-}\phi_{lm} + \phi_{lm}\frac{f}{2r} \frac{\partial x^-}{\partial X^-}\right) \frac{Y_{lm}}{r}
\rightarrow \sum_{lm} \partial_{X^-}\phi_{lm}\frac{Y_{lm}}{R} + \frac{\phi_{J}}{2R}e^{x^+/2R} \ ,
}
but  this is also finite near the horizon.

In summary, we find that there are explicit factors in each of the contributions  to $\phi_J$, which cancel the potentially divergent behavior at the horizon, $x^-\rightarrow\infty$.  As a result, for sufficiently regular $J$, the outward flux $T_{--}$ near the horizon is finite, and the configuration is nonviolent to infalling observers.\foot{Note that violence to infalling observers is relative -- even Hawking radiation is violent, for a sufficiently small black hole.   But  effects scaling to zero as a power of $R$ will be taken to be nonviolent.}  Regularity of $T_{+-}$ can likewise be checked.

Note that one obtains finite stress tensor near the horizon even though a generic $J$ of the form \jexp\ is {\it singular} at the horizon.  To see this, note that 
\eqn\Jsing{e^{-i\omega t} = \left({X^+\over -X^-}\right)^{-i\omega R}\ .}
Thus, $\partial_{X^-}e^{-i\omega t}$ is divergent at the horizon, $X^-=0$.  This behavior may be improved if $j_{\omega lm}(r)$ are chosen so that $J$ vanishes at the horizon, say as a power $(X^-)^p$, though even then \Jsing\ shows that the source is singular.  While such singular but simple sources are nonetheless useful for illustrating the general results of couplings \Jcoup, an additional condition of regularity in the Kruskal coordinates $X^\pm$ may be imposed.  Of course, as explained in section two, these classical sources are merely parameterizations of the effects that arise from the couplings \waveH, \genact\ between the modes in the black hole atmosphere and the internal black hole states.  These are likewise expected to be regular.

An alternate way to characterize the absence of violence at the horizon is in terms of a condition on the state that is created by the nonlocal interactions.  In particular, we can write a ``no-firewall condition" as
\eqn\nofire{a_i|J\rangle \simeq0\ }
(with obvious generalization to states created by the more basic interactions \genact)
where $a_i$ is any annihilation operator corresponding to a Kruskal mode that an infalling observer would see as a high-energy mode when crossing the horizon.

\newsec{Examples and magnitudes}

To understand the size of effects due to effective sources, consider the simple illustrative example
\eqn\Jexamp{J(x) = \sum_{lm} j^0_{lm} \theta(2R-r) e^{-i\omega_{lm} t} \frac{Y_{lm}(\Omega)}{r} +{\rm c.c.}\ ,}
where the $j^0_{lm}$ are constants; the step function cuts the source off at $r=2R$.  The resulting asymptotic flux is given by \Tmm, \Erad, with coefficients $\langle \vectr{u}^f_{\omega l},j_{\omega lm}\rangle$ given by \defbracket\ and modes as pictured as in   \AsympModeDiag.  The mode $\vectr{u}^f_{\omega l}$ in the range $r<2R$ has size governed by the transmission factor $T_{\omega l}$.  For $R\omega\ll l$, this factor is very small; we return to this case shortly.  For $R\omega \roughly>l$ the potential barrier has much less effect, $|T_{\omega l}|\sim 1$.  

To make order-of-magnitude estimates at small $l$, we thus simply approximate the potential as vanishing, and so take $T_{\omega l}= 1$.  Then,
\eqn\approxprod{\langle \vectr{u}^f_{\omega l},j_{\omega lm}\rangle \sim  {j^0_{lm} } {e^{-i\omega_{lm} R}\over -i\omega_{lm}} 2\pi\delta(\omega-\omega_{lm})\ .}
From \Erad, this corresponds to a total radiated energy per unit time
\eqn\nopotrad{ {dE\over dt} \sim \left( {j^0_{lm} \over \omega_{lm}}\right)^2 \ .}

\subsec{Outgoing flux: energy and information}

As described previously, the source $J$ is really a placeholder for the more complicated interactions responsible for transferring and emitting quantum information from the black hole.  In order for black hole evaporation to be unitary, a basic benchmark rate for such transfer is one qubit  emitted per time R, corresponding to the rate of emission of Hawking quanta,
\eqn\Ebench{{1\over T_H}{dE\over dt}\Big|_{\rm bench} \sim {1\over R}\ ,}
where $T_H$ is the Hawking temperature.
Thus, excitations are created with sufficient bandwidth to carry the needed information if
\eqn\jtypg{j^0_{lm} \sim {\omega_{lm}\over R}}
for the relevant modes.  In particular, note that if quanta are emitted with $\omega_{lm}$ appreciably different from $1/R$, but with the same energy flux, the rate of emission is $dE/(\omega_{lm}dt)$ but each quantum carries $\omega_{lm}R$ times more entropy in timing information, so the rate of information transfer is essentially unchanged.

Specifically, suppose as an example that $\omega_{lm}\sim 1/R$.  Then only a few of the lowest-$l$ modes have significant transmission, and with 
\eqn\jtyp{j^0_{lm}\sim 1/R^2\ ,}
 they can carry enough information to restore unitarity.  
If we suppose that interactions of size \jtyp\ are present even for modes with $l\gg 1$ and frequency $\sim 1/R$, they have very little effect on the energy and information that can be transmitted to infinity.  Indeed, through $\langle \vectr{u}^f_{\omega l},j_{\omega lm}\rangle$, the flux \Erad\ in such high-$l$ modes will be suppressed by an extra factor $|T_{\omega l}|^2$ relative to \nopotrad; this is easily seen from \AsympModeDiag\ and the assumption that $j_{\omega lm}$ is insignificant except near the left side of the barrier.
For $R\omega \ll l$, the transmission factors have approximate size\refs{\Star}\foot{This expression is only valid for $R\omega \ll 1$, but WKB gives a comparably small estimate of $\left[\frac{e}{8} \frac{R\omega}{\sqrt{l(l+1)}} \right]^{\sqrt{l(l+1)}} e^{\frac{3\pi}{2}R\omega} \left[1+ \O\left(\frac{R^2\omega^2}{\sqrt{l(l+1)}}\right) \right] $, valid for $R\omega \gg 1/4$ -- see appendix B.  The two estimates approximately match at $R\omega\sim 1$.}
\eqn\transapprox{|T\wl| \sim 2(R\omega)^{l+1} \frac{l!^2}{(2l)!(2l+1)!!} \sqrt{ \prod^l_{n=1} \left[1+\left(\frac{2R\omega}{n}\right)^2 \right] } \ .}
Using Stirling's approximation and ignoring the square root,\foot{The square root is bounded from above by $\sqrt{\frac{\sinh(2\pi R\omega)}{2\pi R\omega}}$.} these are approximately
\eqn\transapproxthree{|T\wl| \sim R\omega \sqrt{\frac{\pi }{2l}} \left(\frac{e}{8}\right)^{l} \left(\frac{R\omega}{l}\right)^{l}}
and they thus give contributions to \nopotrad\ suppressed by a large power of $R\omega/l$.  

For a somewhat different example, suppose that 
\eqn\jspec{\langle \vectr{u}^f_{\omega l},j_{\omega lm}\rangle= j(\omega) T_{\omega l} \ ,}
with $j(\omega)$ independent of $l$ and $m$.  In this case, the radiated energy \Erad\ can be written in terms of the absorption cross section at frequency $\omega$,
\eqn\abscross{ \sigma_\text{abs}(\omega) = \frac{\pi}{\omega^2} \sum^\infty_{l=0} (2l+1)|T\wl|^2 \ .}
Specifically,
\eqn\Erads{E= \int {d\omega \over 4\pi^2} \omega^2 |j(\omega)|^2\sigma_\text{abs}(\omega)\ .}
The Hawking flux is of the same form, with the replacement $|j(\omega)|^2\rightarrow 4\pi \omega \delta(0)/(e^{\omega/T_H} -1)$.
For $R\omega \roughly> \hf$\refs{\Deca}
\eqn\crosshigh{ \sigma \sim \frac{27\pi R^2}{4} \left[ 1 - 8\pi e^{-\pi} sinc \left(\sqrt{27}\pi R\omega \right) \right] \,}
and for $R\omega \roughly< \hf$ \Das,
\eqn\crosslow{ \sigma \sim 4\pi R^2\ .}
Modes with $l\gg R\omega$ again make a relatively small contribution.

\subsec{Ingoing flux, and $R\rightarrow\infty$ correspondence}

Sources like we have described also contribute to an {\it ingoing} radiation flux raining down on observers just outside the horizon,\foot{We thank  R. Bousso for discussions on this point.} described by \Eabs.  Inspection of  \AsympModeDiag\ shows that, taking the representative example \Jexamp, this flux has, for each $l,m$, a similar magnitude to \nopotrad, with {\it no} suppression from the transmission factor $T_{\omega l}$.   This corresponds to an energy density ${\cal E}$ per mode of size $(j^0_{lm}/\omega_{lm}R)^2$, or, in the example $\omega_{lm}\sim 1/R$, with rate from \jtyp, ${\cal E} \sim 1/R^4$ per mode -- the rain is red, in the large-$R$ limit.  Again, as an example, if interactions are present for all $l\leq l_{max}$, the total resulting local energy density near the black hole is of size
\eqn\Eloc{{\cal E} \sim {l_{max}^2 \over R^4}\ .}

This result is important in order to derive a correct correspondence limit for the nonlocal mechanics responsible for the information transfer.  Specifically, we might expect that effects that depart from the LQFT description should vanish in the $R\rightarrow\infty$ limit, since this limit is conventionally viewed as yielding flat space with the black hole exterior corresponding to Rindler space.  For $l_{max} \sim R^k$ with $k<2$, the local energy density \Eloc\ vanishes in this limit.  In particular, note that the maximal mining rate\refs{\Brown} (for more on mining, see below) corresponds to introducing $\sim R$ cosmic strings, and a benchmark for this is 
\eqn\lmax{l_{\max}\sim \sqrt R\ .}
The resulting\NVNLtwo\ extra energy density from \Eloc\ is then\ $\sim 1/R^3$.  Correspondingly, both $T_{--}$ and $T_{++}$ scale to zero as $R\rightarrow\infty$.

It is true that an accelerated observer hovering just outside the horizon sees a blue-shifted version of the energy density \Eloc; specifically, the transformation of the stress tensor to orthonormal coordinates for an observer at $r_0$ gives an energy density of size
\eqn\eboost{\bar{\cal E} \sim {l_{max}^2 \over f(r_0)R^4}\ .}
However, such an observer has proper acceleration $a$, and experiences an Unruh temperature $T_H/\sqrt{f(r_0)}=a/(2\pi)$, with a corresponding energy density\refs{\UnWamine}
\eqn\Undens{\bar{\cal E}_{\rm Un} \sim {1\over f^2(r_0) R^4}\sim a^4\ .}
Thus, in the large-R limit, the size of \eboost\ relative to this characteristic energy density is
\eqn\Erel{\bar{\cal E} \sim {l_{max}^2 \over R^2a^2} \bar{\cal E}_{\rm Un}\ .}
For $l_{max}\ll R$, as in \lmax, and $R\rightarrow\infty$ with $a$ fixed, this contribution is thus negligible by comparison to the effects of the Unruh radiation.

\subsec{Mining and avoiding overfull black holes}

The phenomenon of black hole mining\refs{\UnWamine,\Mining} poses a challenge\refs{\AMPS} to scenarios for unitary black hole evolution, since it allows a black hole to shrink faster than found by Hawking.  In particular, suppose that a black hole has reached a time where the entropy of its radiation equals that describing the number of its internal states; if the latter is $S_{BH}$ this is the Page time\refs{\PageT}.\foot{As described in \refs{\Models\BHQIUE-\GiSh, \NVNL, \NVNLtwo}, the interactions describing information transfer from the black hole (as necessary to restore unitarity) typically imply extra flux and thus\NLSM\ $S_{bh}<S_{BH}$, where $S_{bh}$ is the actual black hole entropy and $S_{BH}$ is the Bekenstein-Hawking entropy, making the corresponding time earlier than the Page time.}  If a mining apparatus is introduced -- a very concrete example is a cosmic string -- the resulting enhancement of the black hole evaporation suggests the possibility of arriving at the inconsistent situation where the entropy of the black hole is smaller than its entanglement entropy with the outgoing radiation; we refer to this as an ``overfull" black hole\refs{\NVNLtwo}.  Of course, what this would really mean, in a quantum mechanical scenario, is that the black hole has more than the expected number of internal states; the final outcome, once the black hole finishes evaporating, would be a Planck-scale remnant, with the resulting inconsistencies\refs{\tHooftRE,\Preskill,\WABHIP,\Susstrouble}.  To avoid this, we expect that, in a consistent scenario, the flux of quantum information out of the black hole should increase commensurately with the increased rate of black hole decay due to mining.

The presence of interactions modeled by sources like those described earlier in this section directly addresses this problem.  Mining corresponds to introducing an additional channel for Hawking radiation to flow out of the black hole.  In the concrete example with a cosmic string, it changes the spectrum of the theory such that there are additional modes whose potential barriers to escaping the black hole are suppressed.  
If there are couplings of the form \Jcoup\ (or more generally, \waveH, \genact) to all such fields that can be mined, and these include in particular the higher-$l$ couplings described above, then opening the extra channel also allows an additional flux of information-bearing excitations created by the source $J$.  In particular, couplings with strengths corresponding to effective sources of size \jtypg\ are parametrically large enough to yield sufficient information transfer, to match the enhanced decay rate of the black hole.  Thus the presence of such couplings gives an in-principle way to avoid the potential problem of overfull black holes resulting from mining.  These couplings to higher-$l$ modes provide a straightforward mechanism to enhance information flow precisely when mining is performed.  This at least partially addresses the ``implausible conspiracy" objections of \AMPSS.

Note also that higher-$l$ interactions like we have described only create appreciable excitation of outgoing modes when a mining channel is opened, {\it e.g.} by introducing a cosmic string.  This may be relevant to discussions\refs{\Bousso} that suggest a special role for ``mineable modes."  Before the mining apparatus is introduced, such modes are not excited and play no obvious special role in the dynamics; in particular, they do not ``carry" the extra quantum information that escapes once mining actually does take place.

It also can be noted that the methods of this paper provide a way to evaluate 
putative scenarios involving manipulation of mined energy/information\AMPS.  Specifically, such manipulations are described, in LQFT, in terms of interactions of the form \effsource, which parameterize the interaction between an experimental apparatus (``external source") and the modes being manipulated.  This provides a means to assess the considerable inherent limitations of such scenarios.

\newsec{Generalizations, extra flux, correspondence, and causality}

While explicit calculations have been performed using an effective source of the form \Jcoup, we stress that this merely serves to illustrate some basic features of the possible information transfer from a black hole.  Again, we expect that this transfer could arise in a  more fundamental description of quantum gravity, which may well not be based on a fundamental spacetime picture.  We do expect that a spacetime picture gives a good {\it approximate} description of a large black hole, for many purposes.  However, transfer of information from the black hole states to excitations that escape to infinity is not described by LQFT.  We may attempt to parameterize it, as a departure from the LQFT dynamics, in terms of interactions of the form \genact.  Then, for the purposes of considering the effects of such interactions on the region exterior to the horizon, we make a further approximation of replacing the interactions by effective sources of the general form \effsource.  

In a complete description of the black hole dynamics, we might expect couplings of such interactions to other operators in the theory, which are more general than those to the fundamental field operators
 in \Jcoup\ (indeed, care is needed to enforce charge conservation for couplings of the latter form).  As noted, a specific and potentially interesting example, given the universal nature of gravitational phenomena, is a coupling to the stress tensor.  A coupling of the form $J^{\mu\nu}T_{\mu\nu}$  would excite modes in all fields.  Indeed, one way to regard the Hawking radiation is as induced from such a coupling between the  non-trivial metric of the black hole, and the stress tensor.  If additional such couplings are present and responsible for the information transfer from the black hole, we may even think of them as analogous to couplings to extra fluctuations of the metric, {\it e.g.} reminiscent of horizon fluctuations.  We expect important features of such couplings to be represented by the behavior of the $J\phi$ couplings we have investigated.  These in particular include the possibility of transmitting, via such couplings, information from the black hole states, at a sufficient rate, without producing singular behavior at the horizon.  

An important point\refs{\BHQIUE,\GiSh,\NVNL,\NVNLtwo} is that generically such couplings produce extra energy flux, beyond that of Hawking, increasing the black hole disintegration rate.  Specifically, the change in the asymptotic flux for  our present example \Jcoup\ is, from \TinJ,
\eqn\deltaT{T_{--}[\phi_J]= (\partial_-\Phi_J)^2\ .}
Such an increased decay rate has important consequences for the statistical mechanics and thermodynamics of black holes\refs{\NLSM}, and in particular indicates a smaller number  of black hole states, with corresponding entropy $S_{bh}$, than given by the Bekenstein-Hawking entropy $S_{BH}$.  A question is whether this conclusion can be avoided, due to special such couplings that do not produce extra flux\refs{\SGWIP}.

A key question, in pursuing a more basic description of the quantum physics incorporating gravity, is that of {\it correspondence}\SGnonpert: specifically, if such mechanics departs from LQFT, it should be well-approximated by LQFT in appropriate limits, including, {\it e.g.}, regimes probed so far by experiment.  For a black hole of size $R$, there are at least two such limits of interest.  

In the first, we consider phenomena  at {\it large} distance from the black hole.  For these, we might anticipate that LQFT gives a good description, as long as we don't for example consider states where strong gravitational effects become relevant to longer scales than $R$.  This in particular motivates the assumption that quantum information transfer from the black hole involves  effects departing from LQFT on scales of size $R$, but not at much larger distances -- in contrast to other proposals.  The latter include proposals with delocalization on enormous scales, such as 
$A=R_B$\refs{\PaRa\VerErr-\ARB} or ER=EPR\refs{\MaSu}.  If  departures from standard locality are only operative on scales $R$, this also indicates how the new effects could  contribute to virtual processes, without leading to larger-scale violations of locality which could be problematic for causality.  Specifically, nonlocalities on scale $R$ do not necessarily imply violation of causality at scales large as compared to the black hole \refs{\NLvC}, providing a way to avoid possible paradoxes due to such real or virtual black hole effects.

In a second such limit  we investigate the vicinity of a large black hole, on scales {\it small} as compared to the black hole.  Here, in classical gravity the equivalence principle would tell us that  a small region near the  black hole is only distinguishable from flat space if we measure effects sensitive to the scale $R$, such as tidal effects.  If the new mechanics are not based on a classical geometrical description, the correct formulation of the equivalence principle is not clear though it may arise from a deeper symmetry principle of the more basic theory.  This means that we do not necessarily expect its classical formulation to hold as an exact statement in quantum gravity.  However, correspondence does suggest that departures from LQFT should likewise vanish parametrically in $R$ for smaller-scale observations near a large black hole -- in contrast to assertions of \refs{\AMPS,\AMPSS} and to expected properties of other scenarios \refs{\Fuzzref}.  We have shown, in section four, that it is possible to introduce interactions with sufficient information carrying capacity to transfer the necessary quantum information, and which also have this property of scaling away in the large-R limit.

Thus, scenarios such as those of \refs{\AMPS} and \refs{\Fuzzref} make the would-be horizon a special -- and likely violent -- place, implying major departure from the equivalence principle, and also calling into question derivation of the Hawking radiation and black hole thermodynamics.
In a nonviolent scenario the deviations from field theory evolution in a semiclassical background only lead to departure from the equivalence principle which make the black hole {\it atmosphere} a special place.  Moreover, the departure is only through ``dilute" effects that scale away in the limit of large black holes.  If this picture is correct,  the equivalence principle as currently formulated remains true in an approximate sense -- as might be expected of a statement about classical spacetime.

\bigskip\bigskip\centerline{{\bf Acknowledgments}}\nobreak

We thank A. Almheiri, R. Bousso, J. Hartle, and J. Santos  for helpful conversations.  This work  was supported in part by the Department of Energy under Contract DE-FG02-91ER40618 and by a Simons Foundation Fellowship, 229624, to SBG.

\appendix{A}{Time Ordering}

For operators whose commutator is central, a time-ordered product like \Jstate\ can be reexpressed  without time ordering.
Specifically, using 
\eqn\expcomm{e^{A_1}e^{A_2}= e^{\frac{1}{2} [A_1,A_2]} e^{A_1 + A_2}\ ,}
%
a time-ordered product can be rewritten
\eqn\contlim{ T e^{\int_{-\infty}^t A(t') dt'} = e^{\frac{1}{2} \int^t_{-\infty}dt' \int^{t'}_{-\infty}dt'' [A(t'),A(t'')]} e^{\int_{-\infty}^t A(t') dt'}\ . }
By assumption of centrality, the extra factor is a complex number; for antihermitan $A$, it is a pure phase. 
%

\appendix{B}{WKB estimate of gray body factors}
Consider a solution of \defu\ with $\omega$ below the barrier given by $V_l$, eq. \defV.  According to the WKB approximation, the transmission coefficient is
\eqn\Twl{ |T\wl| \simeq e^{-\I}\ ,}
with $I$ being the integral between the turning points ${r_{*-}}$ and ${r_{*+}}$,
\eqn\integral{ \I \equiv \int^{r_{*+}}_{r_{*-}} \sqrt{V_l - \omega^2} d\rs\ .}
For large $l$, the $R/r^3$ term in \defV\ is negligible, and $V_l$ can be approximated by
\eqn\newpot{ \tilde{V}_l \equiv f(r) \left[ \frac{l(l+1)}{r^2} \right] \ .}
Note that $\tilde{V}_{l} < V_l < \tilde{V}_{\sqrt{(l+\hf)^2 + 1} - \hf} < \tilde{V}_{l+\frac{1}{2l}}$, which is a tight bound for moderately sized $l$.  Similar considerations apply for the deformed turning points.  These bounds imply that the transmission coefficients calculated with the actual potential $V_l$ can be bounded by those of the modified potential with slightly different $l$:  $|T_{\omega, l+\frac{1}{2l}}|_{\tilde{V}} < |T\wl|_{V} < |T\wl|_{\tilde{V}}$.

Since we are interested in the regime $\Rw \ll l$, it is natural to define a variable whose size characterizes this limit,
\eqn\defA{A \equiv \frac{\sqrt{l(l+1)}}{\Rw} \gg 1 \ .}
For convenience, also define
\eqn\defB{B \equiv l(l+1) \ .}
Using dimensionless parameters, $\mu \equiv r/R$, the integral $\I$ with potential $\tilde{V}_l$ can be rearranged as
\eqn\intapprox{ \tilde\I = \sqrt{B} \int^{\mu_+}_{\mu_-} \frac{\sqrt{f(r)}}{\mu} \sqrt{1 - \frac{1}{A^2}\frac{\mu^2}{f(r)}} \frac{d\mu}{f(r)} \ .}
Between the two turning points,
\eqn\betturn{ 0 < \frac{1}{A^2}\frac{\mu^2}{f(r)} \leq 1\ , }
which is the regime in which the Taylor series for the square root converges.  The endpoints also converge, though parametrically slower.  Thus,
\eqn\infsum{\tilde\I = -\sqrt{B} \sum_{n=0}^\infty a_n \int^{\mu_+}_{\mu_-} \frac{1}{\mu\sqrt{f(r)}} \left[ \frac{\mu^2}{A^2 f(r)} \right]^n d\mu}
where
\eqn\sumcoeff{a_n = \frac{4^{-n}}{2n-1} \frac{(2n)!}{(n!)^2} \ .}
Due to \betturn, each integral is smaller than the previous.  This fact coupled with the fact that $a_n \sim 1/n^{3/2}$ means that the series does indeed converge if the first integral is finite.  The left and right turning points for the modified potential $\tilde V_l$ are, respectively,
\eqn\mumin{ \mu_- =r_-/R= 1 + \frac{1}{A^2} + \O\left( \frac{1}{A^4} \right) }
\eqn\mumax{ \mu_+ =r_+/R= A - \hf + \O\left( \frac{1}{A} \right) \ .}
The integral for $n=0$ of \infsum\ is
\eqn\intone{ \left. \cosh^{-1} (2\mu -1) \right|^{\mu_+}_{\mu_-} = \ln 4A - \frac{3}{A}+ \O\left( \frac{1}{A^2} \right)\ .}
A closed form expression for the integrals in \infsum\ also exists for each $n>0$, but practically, these terms quickly become unwieldy.  Instead, we find leading-order contributions to them in $1/A$.  These integrals can be written as the difference of the function
\eqn\Fdef{F(\mu)={1\over A^{2n}}\int_a^\mu d\mu { \mu^{2n-1}\over \left(1-{1\over \mu}\right)^{n+1/2}} }
evaluated at $\mu_+$ and $\mu_-$; $a$ is arbitrary.  For the former, we expand the integrand of \Fdef\ in $1/\mu$, and integrate term-by-term, using \mumax, to find
\eqn\Fplus{F(\mu_+)={1\over 2n}  + \frac{1}{2n-1} \frac{1}{A} + \O\left( \frac{1}{A^2} \right)\ .}
For the latter, the expansion is in $\mu-1$, and using \mumin\ gives
\eqn\Fmin{F(\mu_-)=- \frac{2}{2n-1} \frac{1}{A} + \O\left( \frac{1}{A^2} \right)\ .}

Adding all the terms  of \infsum\ that are non zero as $A \rightarrow \infty$ gives 
\eqn\sumone{ \sqrt B \left(\ln4A - \sum_{n=1}^\infty \frac{a_n}{2n} \right)
	= \sqrt B \ln \frac{8A}{e} \ ,}
and the sum of terms at order $1/A$ gives
\eqn\sumtwo{-3 \frac{\sqrt{B}}{A} \left( 1+ \sum_{n=1}^\infty \frac{ a_n}{2n-1}\right)
	= -\frac{3\pi}{2} \frac{\sqrt B}{A} \ .}
Combining these gives an estimate for the transmission factor \Twl, via \intapprox, \infsum:
\eqn\Tsol{
|T\wl| \sim \left[ \frac{e}{8} \frac{\Rw}{\sqrt{l(l+1)}} \right]^{\sqrt{l(l+1)}}
	e^{\frac{3\pi}{2}\Rw} 
	\left[1+\O\left({R^2\omega^2 \over \sqrt{l(l+1)}}\right)\right]\ .
}

To understand when the WKB estimate \Tsol\ is good, note that the change of the potential in a wavelength should be small compared to the inverse squared wavelength,
\eqn\wkbcond{\frac{1}{4} \left|\frac{V'}{\left( V - \omega^2 \right)^{3/2}} \right| \ll 1\ .}
This condition holds asymptotically, where both $V$ and $V'$ approach zero.
 In order for \Tsol\ to be a reasonable estimate of the transmission coefficient, \wkbcond\ should hold inside the classically forbidden region.  There, for large $l$, and $R\omega\ll l$, the condition holds as long as $f\approx 1$.  To check the behavior at the lower end of the potential, note that with $\omega^2\ll V$, \wkbcond\ becomes
\eqn\wkbcondtwo{  \left|\frac{V'}{V^{3/2}}\right| \approx {|1-3f|\over \sqrt f \sqrt{l(l+1)}} \ll 4\ .}
Above the turning point, $r/R > 1+1/A^2$, so $f>1/A^2$.  Then, \wkbcond\ is still satisfied as long as $R\omega \gg 1/4$.  

\listrefs
\end